\documentclass[10pt]{amsart}

\usepackage{amsmath}
\usepackage{amsfonts}
\usepackage{amssymb}
\usepackage{epsfig}
\usepackage{graphicx}

\title[Cash Participation]{Quantifying Student Effort and Class Involvement in the Introductory Higher Education Science Classroom}
\date{Novemeber 6, 2007}
\author[Daniel P. Snowman]{Daniel P. Snowman}
\address{Department of Physical Sciences, Rhode Island College, Providence, RI 02908}
\email{dsnowman@ric.edu}

\begin{document}

 \begin{abstract}
This note details a minimal effort program, Cash Participation, that quantifies class participation while increasing class energy, decreasing student apathy and removing student embraced anonymity.
 \end{abstract}

 \maketitle

\section{Introduction}
The primary purpose of this short note is to describe a program, Cash-Participation, introduced to the author's Introductory Physics and Physical Sciences classes over the past few years at Rhode Island College.  This classroom currency system was designed to provide incentive to the students while simultaneously quantifying class participation for the instructor.  This program has created tremendous excitement and improved student performance in four separate introductory classes.  The report that follows details the program and offers reflections on its success.

The major motivating factors that gave rise to this program remain the core of its focus.  The instructor's desire to:  i) quantify class participation, ii) increase energy and involvement; ideally, to the precipice of intellectual rowdiness, iii) decrease student apathy, iv) remove the anonymity comfort factor, and v) provide immediate rewards and continuous feedback.

\subsection{Institutional and Class Demographics}
Rhode Island College is primarily a commuter institution at which most students work 25+ hours per week, simultaneously fulfilling family obligations while maintaining a heavy academic course load.  Many of these hard-working students are more adept at keeping a job than flourishing academically.  In an attempt to capitalize on their working world experiences a currency was established the classroom.  

The Introduction to Physical Science course at Rhode Island College satisfies a General Education requirement, however, the course is primarily populated by Elementary Education majors, for whom the course is required and of which they are, by and large, petrified.  The Introductory Physics sequence, targeted with this program, is populated primarily with biology, chemistry and mathematics majors, respectively.

\section{Background}
Several studies, (Weisz, 1990), (Rau \& Sherman Heyl, 1990), (Kember \& Gow, 1994) support the intuition of virtually every professor that has stood behind the lectern; class participation best facilitates and expedites student mastery of the material.  For most teachers these studies reveal the obvious.  Class participation directly benefits the students and creates a much more dynamic atmosphere in the classroom.

A study by Fritschner (1995) focused on differences in student and faculty definitions and/or interpretations of class participation.  Previous studies, Karp and Yoels (1976) and Howard, Short, and Clark (1996), have revealed that the use of student surveys, as employed by Crawford \& MacLeod (1990) and Heller, Puff, \& Mills (1985), to gauge self-perceived student participation, can often be quite inaccurate.  Student`s perception of their level of participation is often quite different from the actual level of participation.  

Fritschner observed fourteen introductory classes, six 200-level classes, and twelve upper-division classes as class participation was monitored and classified.  The methodology of Howard, Clark, and Short (1996) was used to classify the type of interaction:  (1) instructor initiated, (2) student initiated, (3) direct question to a student, and (4) offhand comment by student.  This study found, in 344 observed class sessions, 28\% of those participated in class.  At the introductory level a mere 8\% of those in attendance made two or more comments during the class and these students accounted for 75\% of all verbal participation.

Fritschner (1995) also explored the differences that exist between student views of participation and faculty views of participation.  In all interviews with students Fritschner found that each considered class participation to be essential to the learning process.  Although ``talkers'' and quiet students both agree class participation is important, it seems that social dynamics lead to a pattern of participation whereby only certain students participate and it is expected by the nontalkers.  Karp and Yoels (1976) introduced the concept of the ``rate buster'', one who raises instructor expectations for the rest of the class.

The reader should also see Fassinger`s (1995) study probing the nature of college classroom interactions from a social psychological perspective.  That is, the role of peer expectations and norms was explored as it relates to classroom interactions.  Fassinger claims that the impact of the professor upon the classroom dynamic has been overestimated and that student silence or involvement is largely a matter of sociological group dynamic beyond control.  The program presented in this paper rebuts this claim as it describes the author`s experiences as the classroom dynamic is unashamedly manipulated to maximize student participation.

\section{The Currency}
Obscure, discarded casino playing cards are employed as the classroom currency.  With this faux currency, 2`s are worth \$200, 3`s are worth \$300, etc up to the face cards worth \$1,000.  Logistically, playing cards allow for a smooth and seemless distribution of the faux cash throughout the meeting time and the obscure nature of the cards is an attempt to avoid and minimize any temptation to 'counterfeit'.

\section{Opportunity}
Cash is earned in a variety of ways.  First – Payday!  Students receive a paycheck of \$1,000 promptly at the start of class.  Tardy students do not recieve a paycheck.    Second, three different rounds of opportunity are presented to the class throughout each meeting:  Opportunity, Double-Opportunity, and Triple-Opportunity.  Each round presents the student with an opportunity to earn additional cash.  The first round of Opportunity allows students to earn between \$200-\$400 in exchange for a contribution (i.e. answering a question, presenting a homework problem, offering an insight, asking a good question, etc.).  The round of Double-Opportunity allows students to earn \$500-\$700 per contribution, and the round of Triple-Opportunity \$800-\$1,000.  Throughout each round of Opportunity, the instructor is circulating throughout the class immediately rewarding student contribution with random cash-draws for that particular round, while continually peddling Opportunity.  The randomness with each draw keeps them excited with anticipation and has minimal affect on their final cash total.  

Cash-Participation offers a wide range of opportunities that will allow all sorts of students of varying strengths and confidences to participate.  Overtly shy students are given the opportunity to erase the board, prepare a template for class data to be recorded, setup and/or takedown equipment, etc.  

Attempts are made to avoid establishing `patterns of participation`, as first identified by Karp and Yoels (1976).  That is, a group of "talkers" is established and the nonparticipating students grow to expect and enforce this pattern.

\section{Cash-Out}
At the end of each day, students tally their earnings, giving some students much needed practice with remedial arithmetic, and cash-out with pats-on-the-back or mild admonishments of you-can-do-better.  Cashing out each student requires only a few seconds and provides valuable daily feedback and face time with each student.  The continual feedback serves to avoid these massive differences in perceived and actual levels of participation as investigated by Fritschner (1995), see section 2.

The cash-out process is very manageable for a class of 24 students.  Each student`s day of hard work is recorded on a Cash-Board, an MS-Excel spreadsheet, available for all to view throughout the semester sans anonymity.  At the end of each day, the Cash-Board is regenerated with great anticipation.  It was the norm for approximately one-third to one-half of the class to linger until all earnings were recorded, the Cash-Board regenerated, class rank reestablished and earnings posted.  The Cash-Board is used directly to quantify class-participation grades used in calculating final averages.  

\section{Results and Reflections}

\subsection{Daily Earnings Goal}Students are given an average cash earnings target, based on the earnings of a fictitious student that participates, on average, two times per class.  It is exciting to report that in four classes, consisting of a total of ninety-six students, approximately seventy-five percent of the students participated two or more times.  This is an extraordinary result, especially when compared to Fritschner`s (1995) study, see section 2, that found only 8\% of students participated two or more times.

\subsection{Exam Improvement}The relative success of this program has been measured by comparing exam averages in eight classes with and eight classes without Cash Participation.  This analysis examined the performance of complementary sections of Introductory Physics I (2 sections), Introductory Physics II (2 sections) and Introduction to Physical Science (4 sections), corresponding to a total eight classes and 196 students.  Discretion was used in this comparison and attempts were made to compare those sections of courses that used similar pedagogical approaches and exams of similar difficulty, etc.  Table I depicts the exam averages in the various courses with and without Cash Participation.

\begin{tabular}{|p{2in}|p{1in}|p{1in}|}
\hline \textbf{Course} & \textbf{Exam Average without Cash Participation} & \textbf{Exam Average with Cash Participation}\\
\hline {Intro Physics I - Sec I} & {64.0}  & {NA} \\
\hline Intro Physics I - Sec II   & 66.0 & NA \\ 
\hline  Intro Physics I - Sec III & NA & 81.0 \\ 
\hline  Intro Physics I - Sec IV & NA & 79.0 \\ 
\hline  Intro Physics II - Sec I & 64.5 & NA \\ 
\hline Intro Physics II - Sec II & 60.3 & NA \\ 
\hline Intro Physics II  - Sec III & NA & 78.8 \\ 
\hline  Intro Physics II - Sec IV & NA & 77.6 \\ 
\hline Intro Physical Science - Sec I & 71.0 & NA \\ 
\hline Intro Physical Science - Sec II & 68.4 & NA \\ 
\hline Intro Physical Science - Sec III & 71.0 & NA \\ 
\hline Intro Physical Science - Sec IV & 74 & NA \\ 
\hline  Intro Physical Science - Sec V &  NA & 75.6 \\ 
\hline Intro Physical Science - Sec VI &  NA & 76.4 \\ 
\hline Intro Physical Science - Sec VII &  NA & 77.6 \\ 
\hline Intro Physical Science - Sec VIII &  NA & 72.3 \\ 
\hline 
\end{tabular}

The Introductory Physics sequence showed remarkable improvement with the exam average jumping from 63.7 to 79.1, corresponding to an increase of 24\%, when grouping Introductory Physics I \& II together for a total of four classes (96 students).  The Introduction to Physical Science course showed a more modest, yet still respectable rise in exam average from 71.1 to 75.5, corresponding to an improvement of 6.2\%, when grouping all four sections together (96 students).

Ultimately, the reasons for these increases are debatable and may be due to any number of factors.  This author is of the opinion, however, that this program resonates especially well with the more reticent students in each class (i.e. the majority) and thus neither these results, nor this pedagogical approach, should be summarily dismissed as a trivialization of education.

 \subsection{Preparedness and Hardwork}In general, students showed to class more prepared with the realization that hard work is guaranteed to be reflected in the final grade.  Students, unlike their instructor, often lack the confidence that hard work translates into improved performance.

 \subsection{Efficiency}The system has allowed the author of this note to maintain a higher level of efficiency with the material coverage.  This is a direct result of a greater percentage of students seeking to participate as they show to class more prepared, thus, ready to engage in the discussion of the day.

 \subsection{Challenges}One problem faced with Cash-Participation has been the number of students seeking to become involved at each Opportunity.  This is a ``problem'', however, that this author will gladly confront and, of course, any conscientious instructor will be sure to provide equal opportunity for each student.

 \subsection{Quantifying Involvement}Traditionally, when class participation was a component of the final grade, the author of this paper would typically weight it at ten percent of the final grade.  However, continually noting the level at which a student is participating is not at all effective and ripe with inaccuracy.  Undoubtedly, the awarding of this component of the grade would be related to attendance and/or general, overall impressions students had formed over the course of the semester.  Obviously, this is not a very scientific process nor is it quantitative in the least.  A better method was needed and the Cash-Participation program presented in this paper is a vast improvement, though far from perfect.  With the ability to quantify student involvement and discriminate between different levels of participation, the class participation component of the final grade was increased to twenty percent.

\subsection{Personal Approach}Finally, for more intimate gatherings of students, twenty-four or fewer, Cash-Participation offers a much more personal approach than the Peer Instruction (Mazur) with Class Response System, see James (2006).  While the Class Response System is effective in the larger classroom, the author of this paper prefers Cash-Participation since it simultaneously maximizes feedback, face-time and participation while lifting the veil of anonymity many students eagerly embrace, particularly at the introductory level.

\section{Conclusions}
Cash-Participation has been a very effective program in energizing the author`s classroom while simultaneously engaging the student and quantifying class participation.  The results have been extraordinary and the program has been a tremendous success.

\section{Acknowledgements}
The author extends thanks to Rhode Island College students populating Physical Science 103 and Introductory Physics 101 and 102 students that participated with Cash-Participation.

\end{document}